\documentclass{article}
\usepackage{ijcai21}

\usepackage{times}

\usepackage{soul}
\usepackage{url}
\usepackage[draft]{hyperref}
\usepackage[utf8]{inputenc}
\usepackage[small]{caption}
\usepackage{graphicx}
\usepackage{amsmath}
\usepackage{amssymb}
\usepackage{booktabs}
\usepackage{latexsym}
\usepackage{float}
\usepackage{xspace}
\usepackage{algorithm}
\usepackage[noend]{algpseudocode}
\usepackage{epsfig}
\usepackage{multirow}
\usepackage{epstopdf}    
\usepackage{subcaption}
\usepackage{booktabs}
\newtheorem{theorem}{Theorem}
\newtheorem{lemma}{Lemma}
\newtheorem{corollary}{Corollary}

\title{ Decentralised Approach for Multi Agent Path Finding}

\begin{document}

\maketitle

\begin{abstract}
Multi Agent Path Finding (MAPF) requires identification of conflict free paths for agents which could be point-sized or with dimensions. In this paper, we propose an approach for MAPF for spatially-extended agents. These find application in real world problems like Convoy Movement Problem, Train Scheduling etc. Our proposed approach, Decentralised Multi Agent Path Finding (DeMAPF), handles MAPF as a sequence of path-planning and allocation problems which are solved by two sets of agents Travellers and Routers respectively, over multiple iterations. The approach being decentralised allows an agent to solve the problem pertinent to itself, without being aware of other agents in the same set. This allows the agents to be executed on independent machines, thereby leading to scalability to handle large sized problems. We prove, by comparison with other distributed approaches, that the approach leads to a faster convergence to a conflict-free solution, which may be suboptimal, with lesser memory requirement.
\end{abstract}

\section{Introduction}

Multi Agent Path Finding (MAPF) is a well researched problem, wherein multiple agents are scheduled to move from their defined start locations to destinations without conflicting with any others. This requires planning for the best path for multiple agents on a shared transport network. The problem becomes complex when multiple agents attempt to use a common path leading to cascading contentions, which need to be resolved.  MAPF can be mapped to many real-world scenarios, which have been comprehensively captured in~\cite{Ma2016}. There have been some attempts at handling \textit{sized} agents, variously termed as multi-sized agents ~\cite{Harabor2008}, spatially extended agents~\cite{Thomas2015}, train agents~\cite{Atzmon2019} and  large agents~\cite{Li2019}. Sized agents occupy multiple locations at the same instance of time, the time of movement from one location to another becomes durative in nature, similarly, edge transitions also cannot be considered to be instantaneous but become durative in nature. \\
\indent The primary contribution of this paper is the formulation of a decentralised approach to MAPF for spatially extended agents i.e. agents which have a length, like convoys, trains etc. We model every entity, i.e. the moving elements and the network elements (nodes and edges) as agents belonging to two sets, the Travellers and the Routers respectively. The Traveller agents, which are required to move from defined source to destination, generate plans for their own movement. The Router agents manage a network resource (node or edge) each, ensuring two or more Travellers do not conflict on their managed network location. In our algorithm, Decentralised Multi Agent Path Finding (DeMAPF), the problem space is thus divided into a series of planning and allocation problems. Several iterations of negotiations between the two sets of agents allow them to arrive at a conflict-free solution. This approach is decentralised as decisions are made individually by each Router and Traveller; no agent has a complete knowledge of the total moving agents, moreover as an agent only makes a decision for itself hence its compute requirements are significantly lower. Further, in terms of implementation, all the agents can be implemented and executed independently on different compute platforms, thereby allowing for scalability for handling large problem spaces.\\
\indent The paper is organised to cover a brief overview on related work in section~\ref{section_relatedWork}. The problem formulation, and the approach follow in sections~\ref{section_problemDefinition},~\ref{section_approach}. The subsequent sections~\ref{section_theoreticalAnalysis}, ~\ref{section_result}, cover the theoretical and empirical analysis, and we finally conclude with our future work and summarisation in sections~\ref{section_futurework},~\ref{section_conclusion}.\\

\section{Related Work} \label{section_relatedWork}
\indent  Distributed approaches for MAPF using search include, Conflict Based Search~\cite{Sharon2015} which proposed an optimal, complete path planning solution for multiple agents using a two level search on a Constraint Tree by a centralised agent. Notable variations and extensions to the approach include meta-agent based CBS~\cite{Sharon2012} in which agents with multiple conflicts are grouped together,  Improved Conflict Based Search (ICBS)~\cite{Boyarski2015} in which plans were restarted for merged agents and conflicts were prioritised. The Push and Rotate approach~\cite{De2014} deconflict agents on a biconnected graph by executing push and rotate actions, using graph bridges. These approaches assumed agents as point sized entities. Sized-agents were considered in Multi-Constraint Conflict Based Search (MC-CBS)~\cite{Li2019} which planned for agents with definite geometric shapes occupying multiple locations. Multi-Train Conflict Based Search (MT-CBS)~\cite{Atzmon2019} handled agents which occupied multiple locations at the same time instance owing to their length, termed as train agents, but the agent transitions from one location to another were instantaneous in nature. Continuous-Time Conflict Based Search (CCBS)~\cite{Roni2019} was an adaptation of CBS and a customized version of Safe Interval Path Planning (SIPP) for a sound, complete and optimal solution. The awareness of other agents plans was exploited in Extended Conflict Based Search with Awareness (XCBS-A)~\cite{Thomas2018} and XCBS with Local Awareness (XCBS-LA)~\cite{Thomas2020}. In each of these approaches, there existed a central entity which made decision based on the states of all other agents which was known to it. \\	 
\indent Decentralised approaches to MAPF typically divided the spatial area into independent agents and then the area agent would plan for all entities passing through it, Spatially Distributed Multiagent Planner (SDP)~\cite{Botea2014}, \textit{ros-dmapf}~\cite{Poom2019} are approaches along this direction. Another strategy was proposed in DiMAPP~\cite{Niyogi2015} with individual agents first planning their path, then a central planner would assign priorities to the agents. The conflict resolution is then ensured in favour of the higher priority agent.\\
\indent Similar dencentralised approaches have also been adopted in multi agent robot path planning, where agents collaborate with reach other to dynamically find the path to their destination as they are moving. In~\cite{Desaraju2011}, Decentralized Multi-Agent Rapidly-exploring Random Tree (DMA-RRT) is proposed where every robot initially plans a path for itself, which is then shared with other robots and in each iteration, the robot with highest merit, replans its path based on known plans of the others. This approach required all robots to be updated with the planned paths of others. A Graph neural network was proposed in~\cite{Li2019} for propagating local observations for online planning.

\section{Problem Definition} \label{section_problemDefinition}
Multi Agent Path Finding with Spatially Extended agents involves planning a conflict free path for a set of agents on a road-network. The solution to the problem is the combined conflict-free path of all the agents from their respective source locations to their destination locations, minimising a defined objective function like the makespan or cumulative time. \\
\indent In this paper, we propose an approach called Decentralised Multi Agent Path Finding (DeMAPF). We model MAPF as an interaction between two sets of agents, set $\mathcal{A}$ of Travellers and the set $\mathcal{R}$ of the Routers, leading to a conflict-free scheduling. Each Traveller $a_j \in \mathcal{A}$ has a defined length $\mathbb{L}_j$, average speed $\mathbb{S}_j$ and  travels on a road network ${G(V,E)}$. It has to be scheduled from a defined source location $\mathbb{I}_j$, to destination location $\mathbb{F}_j$, where $\mathbb{I}_j, \mathbb{F}_j \in {V}$, starting not before the defined start time $\mathbb{D}_j$. The constraint with respect to starting time of the Traveller makes the problem formulation more consistent to practical problems like Convoy Movement Problem~\cite{Chardaire99,RamKumar2011,Khemani2012}, Train Scheduling~\cite{Bettinelli2017,Wang2018} etc. \\
\indent A Traveller generates a plan $P_j$ which defines the locations which lie on the path from $\mathbb{I}_j$ to $\mathbb{F}_j$ and the time spent by Traveller $a_j$ on each of them. The plan is represented as $P_j = \{l_1(t_{1_e}, t_{1_x}),l_2(t_{2_e}, t_{2_x}),\dots,l_k(t_{k_e}, t_{k_x})\}$, where, $l_i \in $ \textbar$V\cup E$\textbar, $l_i$ and $l_{i+1}$  are adjacent in the $G$, and $l_1 = \mathbb{I}_j$, $l_k = \mathbb{F}_j$. The time of entry onto the location $l_i$ is indicated by $t_{i_e}$, similarly, $t_{i_x}$ indicates the exit time. The time $t_{1_e}$ of starting from the source location $\mathbb{I}_j$, is not before defined start time $\mathbb{D}_j$, i.e. $t_{1_e} >= \mathbb{D}_j$. The duration of time spent on a location $l_i$, $(t_{i_e}, t_{i_x})$, depends on the length  $\mathbb{L}_j$,  the speed $\mathbb{S}_j$ of the Traveller $a_j$, as well as the length $\mathcal{L}_i$,  and speed constraints $\mathcal{S}_i$ of the location and is given by the Equation (~\ref{eq-time-to-cross-location}). \\
\begin{equation} \label{eq-time-to-cross-location}
 (t_{i_e}, t_{i_x}) = ( \mathbb{L}_j+ \mathcal{L}_i)/min(\mathbb{S}_j, \mathcal{S}_i)
\end{equation}
A \textit{wait}, for a duration of $t$ timeunits, may be imposed on a Traveller on a given location $l$. A wait results in the Traveller staying in location $l$, and all trailing locations of its plan, which it has not exited, for the given duration of time. The plan cost is defined by Equation (~\ref{eq-planCost}), where the first term is the cumulative time taken to travel all $k$ locations in its plan, the second term computes the cumulative wait time which may be imposed on the agent. 
 \begin{equation}\label{eq-planCost}
  c(P_j)=\sum_{i=1}^k{\biggl\lceil\frac{\mathcal{L}_i + \mathbb{L}_j }{min(\mathbb{S}_j,\mathcal{S}_i)}\biggr\rceil}+\sum_{r=0}^l{wait_{r}} 
 \end{equation}
 
\indent Several Travellers may occupy the same network resource during their movement. The scheduling on the Travellers on the shared network entity is done by the Router agent. Each Router ${r}_i \in \mathcal{R}$ manages a location $l_i$ of the road network, such that \textbar$ \mathcal{R}$\textbar $=$ \textbar$V\cup E$\textbar. The Router $r_i$, generates an allocation for its managed location $l_i$, and is given by, $A(r_i)=\{a_j(t_{j_e}, t_{j_x}), a_{j+1}(t_{{j+1}_e}, t_{{j+1}_x})\dots \}$. The allocation $A(r_i)$ indicates that location $l_i$, is occupied by a Traveller $a_j$ from time interval $t_{j_e}$ to $t_{j_x}$.  The allocation $A(r_i)$: 
\begin{itemize}
 \item is temporally sequenced, i.e. $ t_{j_e} < t_{{j+1}_e}$.
 \item maintains a minimal separation time $t_{min}$ between any two Travellers. i.e. $(|t_{j_x} -  t_{{j+1}_e})| >= t_{min}$, where  $t_{j_x}$ is the time of exit of the earlier Traveller, and , $t_{{j+1}_e}$ is the entry time of the later Traveller. This implies two or more Travellers can simultaneously occupy the same location maintaining minimal spacing between them.
 \end{itemize}
\indent If $a_j(t_{j_e}, t_{j_x})$ belongs to allocation $A(r_i)$ of Router agent $r_i$, it implies, that there exists a corresponding entry $l_i(t_{i_e}, t_{i_x})$, in plan $P_j$ of Traveller $a_j$, where $l_i$ is the location managed by $r_i$. In this paper hereafter, we will follow the convention that Router $r_i$ manages location $l_i$ of the road network.\\
 \indent The solution $\mathfrak{S}=\{P_1, P_2,\dots,P_{|\mathcal{A}|}\}$, comprises the plans of all Travellers. The solution-cost is given by Equation (~\ref{eq-soluCost}), where $c(P_j)$ is the cost of Traveller plan.  The objective of the problem is that the solution $\mathfrak{S}$ should be conflict-free, i.e. any two Travellers should always maintain a minimum spacing $t_{min}$, between them. Secondly, the cost of the plan of each Traveller, $c(P_j)$ should be minimal.\\
 \begin{equation}\label{eq-soluCost}
  Cost(\mathfrak{S})=\sum_{j=1}^{|\mathcal{A}|}{c(P_j)} 
 \end{equation}
 
\section{Decentralised Multi Agent Path Finding}\label{section_approach}
\indent The DeMAPF is a decentralised formulation, wherein a Traveller communicates with a set of Routers to arrive at its plan. The Traveller is not aware of the other Travellers and makes its plans purely based on responses from the Routers it corresponds with. Similarly with the Router agents. \\ 
\indent The assumptions made in this formulation are:
\begin{enumerate}
\item An agent receives all the messages sent to it simultaneously. 
\item The communication between agents is perfect and no messages are lost. This means that a Router receives messages from a Traveller in the order in which the Traveller transmits it (and vice-versa).
\end{enumerate}
\indent We first define some terms used in our approach and then follow up with detailing the behaviours of the Traveller and the Router agents.
\subsection {Definitions}
\indent \textbf{Definition 1} The time-past-a-point $tpp$, is defined as the time taken by the Traveller to move from the end of one location to the start of the next adjacent location and is a constant for a given Traveller, depending solely on its length $\mathbb{L}_j$ and its speed $\mathbb{S}_j$. It is given by the Equation (~\ref{eq-tpp}). 
\begin{equation}\label{eq-tpp}
 tpp =  \mathbb{L}_j /  \mathbb{S}_j
\end{equation}
\textbf{Definition 2} A proposed-plan $\hat{P}_j$, is formed by the Traveller $a_j$, by temporally ordering the allocations for $a_j$ by $k$ Routers. \\
$\hat{P}_j = \{l_1(t_{1_e}, t_{1_x}),l_2(t_{2_e}, t_{2_x}),\dots,l_k(t_{k_e}, t_{k_x})\}$, given, \\
$A(r_1)=\{\dots a_j(t_{1_e}, t_{1_x})\dots \}$, \\ $A(r_2)=\{\dots a_j(t_{2_e}, t_{2_x})\dots \}$, \\$A(r_k)=\{\dots a_j(t_{k_e}, t_{k_x})\dots \}$ and $t_{i_e}<t_{{i+1}_e}$.\\

\textbf{Definition 3} A proposed-plan $\hat{P}_j$, is \textit{consistent} only if the allocations for the Traveller, given by the Routers are \textit{feasible} and \textit{contiguous}. Let $\hat{P}_j = \{l_1(t_{1_e}, t_{1_x}), l_2(t_{2_e}, t_{2_x}),\dots,l_k(t_{k_e}, t_{i_{k_x}})\}$ be a proposed-plan, then $\hat{P}_j$ is said to be feasible if $l_i$ and $l_{i+1}$ are adjacent locations in $G$. Secondly, if $t_{i_x} - t_{{i+1}_e} = tpp$ then the time-slots are said to be contiguous. \\
\subsection{Traveller Agent}\label{subsection_travellerAgent}
A Traveller Agent generates and maintains all possible plans for itself in a search tree. If a plan has conflicts, then constraints are imposed on the conflicting locations and alternate plans generated. To maintain and search through all possible plans based on different constraints, the Traveller uses the Constraint-Tree (CT) as defined in~\cite{Thomas2015}. Each node of the CT comprises a consistent plan, constraints imposed for the generation of the plan and the cost of the plan of the Traveller. The constraints imposed on the plan are inherited down the branch of CT for generation of newer plans. The functionality of the Traveller is shown in Algorithm~\ref{algo-travellerAgent}.\\
\indent Initially, the root CT node will have no constraints, and the plan $P_j$ is generated assuming that the road network is unoccupied, as shown in Line~\ref{algo-travellerAgent-getInitPlan}. An open-set $O$ is used to maintain a cost-ordered list of unexplored leaf nodes of the CT. The generated root CT-node is added to the open-set, as shown in Line~\ref{algo-travellerAgent-addInitialPlanToOpenSet}. The Traveller then chooses the plan $P_j$ from the least-cost CT-node of the open-set for negotiation with the Routers, shown in Line~\ref{algo-travellerAgent-chooseBestPlan}. For every location $l_i$ in $P_j$, the Traveller formulates a request $req(l_i(t_{i_e}, t_{i_x}))$ for the corresponding managing Routers, as in Line~\ref{algo-travellerAgent-sendReserve}. The request also contains header information with the speed and length of the Traveller. Each Router responds to the request with a proposed allocation $a_j(t_{i_{pe}}, t_{i_{px}})$, as discussed in Section~\ref{subsection_routerAgent}. The proposed allocations are formulated into a temporally ordered proposed-plan $\hat{P}_j$, as in Line~\ref{algo-travellerAgent-getProposed}, such that $\hat{P}_j =\{ l_i (t_{i_{pe}}, t_{i_{px}}), l_{i'}(t_{{i'}_{pe}}, t_{{i'}_{px}})\dots\}$, $t_{i_{pe}} < t_{i'_{pe}}$, and checked for consistency. \\
      
  \begin{algorithm}[h]
	    \caption{\textbf{Traveller Agent}}\label{algo-travellerAgent}
	    \begin{algorithmic}[1]
		    \State Make plan $P_{j}$ using $shortestPath(\mathbb{I}_j, \mathbb{F}_j) $  \label{algo-travellerAgent-getInitPlan}
		    \State Define CT-node $N_r$
		    \State Insert $N_r$ to Open-Set $O$. \label{algo-travellerAgent-addInitialPlanToOpenSet}
		    
		    \While{$!empty(O)$ }
		      \State $P_j \gets N_{least}.Plan \gets least(O)$ \label{algo-travellerAgent-chooseBestPlan}
		      \State Send Request $req(l_i(t_{i_e}, t_{i_x}))$ to Routers for all locations $l_i$ in $P_j$. \label{algo-travellerAgent-sendReserve}
		      \State Proposed plan $\hat{P}_j \gets$ Receive($Allocation(r_i)$)  \label{algo-travellerAgent-getProposed}
		      \If { !Consistent($\hat{P}_j$) } 
		        \State $PlanFound \gets False$ \label{algo-travellerAgent-setPlanNotFound}
			\State $\hat{P}_{rev} \gets make-consistent(\hat{P}_j$) \label{algo-travellerAgent-makeConsistent}
			\State Define $\hat{N}$ with $\hat{P}_{rev}$. 
			\If{$!duplicate(\hat{N})$}
			  \State Insert $\hat{N}$ to $O$ \label{algo-travellerAgent-checkDuplicate}
			\EndIf
			\State $l_i \gets firstInconsistentLocation(\hat{P}_j, P_j)$
			\State Make plan $\bar{P}_j$ using $shortestPath(\mathbb{I}_j, \mathbb{F}_j$, exclude $l_i)$.
			\State Define $\bar{N}$ with $\bar{P}_j$. 
			\If{$!duplicate(\bar{N})$}
			  \State Insert $\bar{N}$ to $O$
			\EndIf  
		    \Else 
		      \State $N_{final} \gets N_{least}$ \label{algo-travellerAgent-setNFinal}
		      \State $PlanFound \gets True$ \label{algo-travellerAgent-setPlanFound}
		     \EndIf
		
		    \EndWhile
		    \If{$PlanFound$}
		      \State Add $N_{final}.Plan$ to $Solution$.\label{algo-travellerAgent-accept}
		    \EndIf
	    \end{algorithmic}
    \end{algorithm}
    
\indent If $\hat{P}_j$ is consistent, $N_{final}$ is set with the currently explored CT-node (Line~\ref{algo-travellerAgent-setNFinal}) and $PlanFound$ is set to True (Line~\ref{algo-travellerAgent-setPlanFound}).\\
\indent If $\hat{P}_j$ is not consistent, the Traveller generates a revised consistent proposed plan,  $\hat{P}_{rev}$, by introducing a \textit{wait} action on the first location where the delay was imposed. The delay is also added to all the locations later in plan. This is done in the $make-consistent$ function in Line~ \ref{algo-travellerAgent-makeConsistent}. If $l_i$ be the first location which deviated from the original plan $P_j$, then the Traveller also generates an alternate plan $\bar{P}_j$ by introducing a constraint of not using $l_i$ for time $t_{i_e}$ to $t_{i_{pe}}$ to find the shortest-path. This enables to find an alternate diversionary route. The CT-nodes $\bar{N}$, $\hat{N}$ so formed are added to the open-set, if no duplicates exists i.e. there exist no other CT-node with the same plan. \\
\indent  The search terminates successfully when the plan is found. The $N_{final}$ is added to the $Solution$, shown in Line~\ref{algo-travellerAgent-accept}. If 
there are no more CT-nodes to be explored and yet the plan is not found, the algorithm terminates unsuccessfully. \\

\subsection{Router Agent}\label{subsection_routerAgent}  
The Router agent manages the allocation of a location based on requests from Traveller agents, such that no two Travellers overtake each other when they occupy the location. A precedence ordering of the incoming Travellers is generated by the Routers. We have defined the ordering in terms of speed and length of the Traveller, however any other characteristic of the Traveller which does not change during the plan, can be taken for the precedence ordering. Further, all Routers have the same ordering for the Travellers, i.e. if Routers $r_1, r_2$ receive requests from Traveller agents $a_1, a_2$,  then the ordering for $a_1$ and $a_2$ will be same for both Routers. Each request from a Traveller is of the form $l_i(t_{i_e}, t_{i_x})$, where $t_{i_e}$ implies the entry time and $t_{i_x}$, the exit time of a Traveller. The Router schedules the incoming requests for the location with constraints that (\textit{i}) a Traveller cannot be scheduled earlier than its requested entry time $t_{i_e}$,  (\textit{ii}) the duration of time requested by the Traveller needs to remain unchanged as it depends on the time taken to traverse the location completely. The Router allocates the time-space of the managed location by ordering the requests in terms of the precedence of the Traveller agents. The Traveller with the higher precedence (larger speed, longer length) is given its requested timeslot as is and the lower precedence Travellers are allocated to other non-conflicting time-slots. \\
\indent In Algorithm~\ref{algo-routerAgent}, the Router agent receives all Requests from Travellers (Line~\ref{algo-routerAgent-receiveRequest}). 
At Line~\ref{algo-routerAgent-orderRequest}, the Router temporally orders all the requests, based on speed and then length of the Traveller. The Router then iterates through the ordered list of requests, and checks for overlap of the requests time-slot. A requested time-slot $(t_{i_e},t_{i_x})$ is said to overlap another time-slot $(t_{j_e},t_{j_x})$ if $(|t_{j_x} -  t_{i_e})| < t_{min}$, or vice-versa. If the requested time-slot does not overlap with the Reserve list, which maintains time-slots which have been proposed for a request earlier in the list, then the requested timeslot is allocated as is and set as a proposal (Lines~\ref{algo-routerAgent-iterate}-~\ref{algo-routerAgent-setTimeSlot}). If there is an over-lap, then the request is allocated the next earliest non-overlapping timeslot, as in Line~\ref{algo-routerAgent-computeNextTimeSlot}. The Reserve list is updated with the proposed time-slots for each request. When all the requests have been allocated, the proposals are sent back to the respective Traveller agents.\\

   \begin{algorithm}
	    \caption{\textbf{Location Agent}}\label{algo-routerAgent}
	    \begin{algorithmic}[1]
		    \State $Reserve \gets \phi$
		    \State $Requests \gets$ Receive($req(l_i(t_{i_e}, t_{i_x}))$) \label{algo-routerAgent-receiveRequest}
%
		    \State $Requests \gets Sort(speed, length)$\label{algo-routerAgent-orderRequest}
		    \While {$!Requests.empty$} \label{algo-routerAgent-iterate}
		    \State $R_i \gets next(Requests)$
			\If {$!In(R_i.TimeSlot, Reserve)$ }
			    \State $p_i \gets R_i.TimeSlot.$\label{algo-routerAgent-setTimeSlot}
			\Else
			   \State $Allocation_i \gets Compute(nextTimeSlot)$\label{algo-routerAgent-computeNextTimeSlot}
			 \EndIf
		    \State Add $Allocation_i$ to $Reserve$. \label{algo-routerAgent-addToReservation}  
		    \EndWhile  
		    \State $Send(Allocation)$.
	    \end{algorithmic}
    \end{algorithm}

\section{Theoretical Analysis}\label{section_theoreticalAnalysis}

\begin{lemma}\label{lemma_costCT}
The cost of plans, for a given Traveller, is non-decreasing down the Traveller's Constraint Tree.\end{lemma}
\textit{Proof}
By induction: The root CT-node, $N_r$ has $N_r.Plan = \{l_1(t_{1_e},t_{1_x},\ldots\}$ where $N_r.Plan$ is the schedule on the shortest path between the source and destination of the Traveller and hence is the least cost node in the Tree.\\
Let $N_e$ be the CT-node being explored, then $N_e.Cost >= N_{\tilde{e}}.Cost$, where $N_{\tilde{e}}$ is the parent of $N_e$. 
Let $\hat{P}_j$ be the proposed-plan formulated against requests on $N_e.Plan$. If $\hat{P}_j$ is not consistent, then let $N_{e1}$ be the child CT-node generated by imposing constraint on the first location $l_i$ which deviates from $N_e.Plan$. As the $N_e.Plan$ was the least cost plan with location $l_i$, any other plan with a constraint on $l_i$ will have equal or more cost, hence $N_{e1}.Cost >= N_e.Cost$. Similarly, $N_{e2}.Plan$ is formed by disseminating the delay of $l_i$ wrt requested allocation, to all locations later in the plan, this will lead to increase in the cost of the plan with respect to the parent node. Hence $N_{e2}.Cost > N_e.Cost$.
\begin{corollary} \label{lemma_leafNodes}
The leaf-nodes of the CT are least-cost unexplored nodes in the CT. 
\end{corollary}

\begin{lemma} \label{lemma_consistent}
A proposed plan, $\hat{P}_j$, is consistent only when the proposals exactly match the requests made by the Traveller. \end{lemma}
\textit{Proof:} The Traveller agent only maintains consistent plans in its search tree, hence $P_j$, which is the least-cost plan from open-set, is consistent. Let $\hat{P}_j$ be a consistent proposed plan responded against $P_j$. A Router $R_i$ only responds to a message from a Traveller, and responds to all the Traveller messages it receives, hence $\hat{P}_j$ only contains the locations which were part of $P_j$, and no new locations are introduced in $\hat{P}_j$. If $l_i(\hat{t}_{i_e}, \hat{t}_{i_x}) \in \hat{P}_j$ be different from the requested time-slot $l_i(t_{i_e},t_{i_x})$. As a Router can only delay a request, it implies that $\hat{t}_{i_e} > t_{i_e}$. This means, that the time-slots by the location lying adjacent to $l_i$ will either not be contiguous and/or may overlap. This is a contradiction to the definition of a consistent plan. \\
\begin{figure*}
	\centering
	\begin{minipage}{0.3\textwidth}
		\includegraphics[scale=0.38]{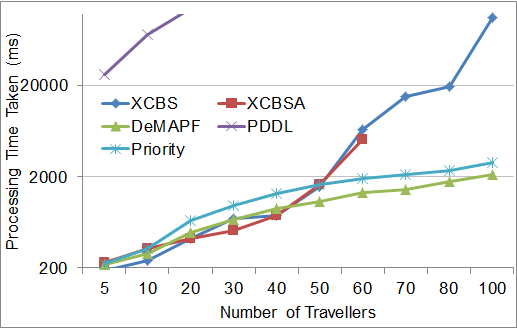}
		\subcaption{Number of Traveller Agents }
		\label{fig:Performance-grid}
	\end{minipage}%
	\begin{minipage}{0.3\textwidth}
		\includegraphics[scale=0.35]{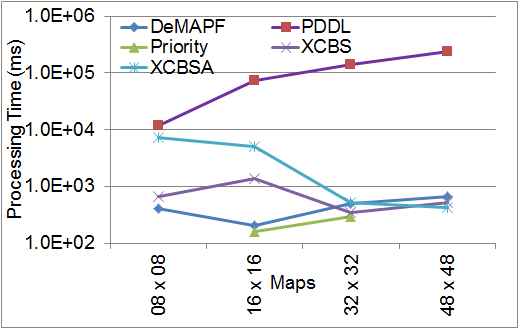}
		\subcaption{Different maps}
		\label{fig:networks}
	\end{minipage}%
	\begin{minipage}{0.3\textwidth}
		\includegraphics[scale=0.38]{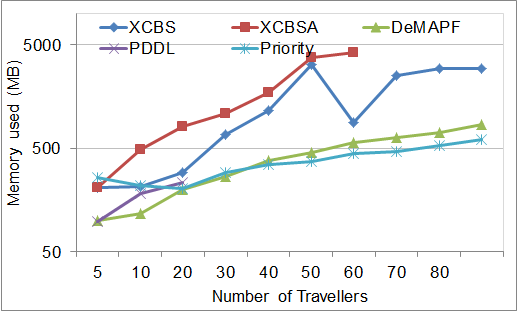}
		\subcaption{Memory Requirement}
		\label{fig:Memory}
	\end{minipage}
	\begin{minipage}{0.3\textwidth}
		\includegraphics[scale=0.3]{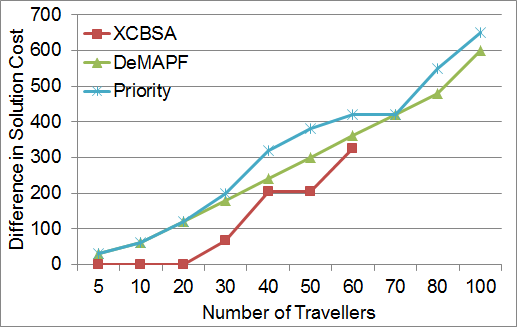}
		\subcaption{Difference in Solution Cost}
		\label{fig:SolutionQuality}
	\end{minipage}%
	\begin{minipage}{0.3\textwidth}
		\includegraphics[scale=0.3]{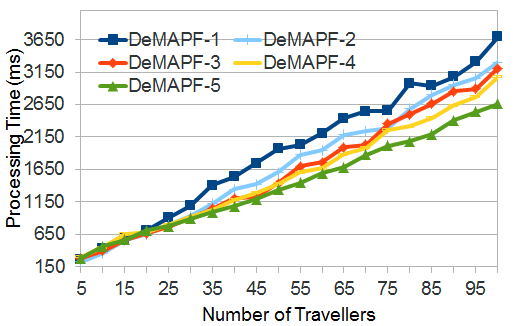}
		\subcaption{Distributivity}
		\label{fig:Distributivity}
	\end{minipage}%
	
	\caption{Evaluation of the performance of DeMAPF with other algorithms}
\end{figure*}

\begin{lemma}\label{lemma_acceptance}
The proposed plan accepted by a Traveller is the least-cost plan for it in its search tree.\end{lemma}
\textit{Proof:} The unexplored leaf-nodes of the Constraint Tree are maintained by the open-set $O$ of the Traveller. From Corollary~\ref{lemma_leafNodes}, the least-cost node of the $O$, $N_l$ is also the least-cost unexplored node of the CT. Let $P_l = N_l.Plan$, be the plan, against which the proposed plan formulated by the Routers be $\hat{P}_j$. If the Traveller accepts $\hat{P}_j$, then from Lemma~\ref{lemma_consistent}, $\hat{P}_j = P_l$, and hence $\hat{P}_j$ is the least-cost plan in its search tree.  \\

\begin{theorem}\label{theorem_agent_terminate}
A Traveller will converge to an acceptable plan in maximum $2^{(k-1)}$ iterations, where $k$ is the number of unique locations traversed by Travellers of higher precedence.\\\end{theorem}
\textit{Proof:} In each iteration of a Traveller's search, a constraint is imposed in the CT with the first location which conflicts with higher precedence Travellers. Each conflict creates two CT-nodes, and the constraints are inherited down the branch of the CT. Hence $k$ unique locations will lead to $k$ levels down a branch of the CT. The worst case scenario leads to the exploration of a full binary tree with $2^{k-1}$ nodes being explored, hence worst case scenario of $2^{(k-1)}$ iterations.\\


\begin{table*}[!hbt]
	\small
	\centering
	\caption{Ranking of the performance of different algorithms on evaluated properties}
	\label{table-summary}
	\begin{tabular}{|c|c|c|c|c|c|c|}
		\hline
		\multicolumn{1}{|c|}{\multirow{2}{*}{Algorithms}} & \multicolumn{6}{c|}{Properties} \\ \cline{2-7} 
		& \multicolumn{1}{|c|}{\multirow{1}{*}{\begin{tabular}[c]{@{}c@{}}Decentralised\end{tabular}}} & \multicolumn{1}{c|}{\multirow{1}{*}{\begin{tabular}[c]{@{}c@{}}Distributed\end{tabular}}} & \multicolumn{1}{c|}{\multirow{1}{*}{\begin{tabular}[c]{@{}c@{}}No of Agents\end{tabular}}} &
		\multicolumn{1}{c|}{\multirow{1}{*}{\begin{tabular}[c]{@{}c@{}}Map size\end{tabular}}} &
		\multicolumn{1}{c|}{\multirow{1}{*}{\begin{tabular}[c]{@{}c@{}}Optimality\end{tabular}}} &
		\multicolumn{1}{c|}{\multirow{1}{*}{\begin{tabular}[c]{@{}c@{}}Memory Rqmt\end{tabular}}}\\
		\hline
		\begin{tabular}[c]{@{}c@{}}PDDL Based\end{tabular}	& No & No & 5 & 5 & - & -\\ \hline
		\begin{tabular}[c]{@{}c@{}}XCBS\end{tabular} 
		& No & Yes & 3  & 3 & 1& 3\\ \hline
		\begin{tabular}[c]{@{}c@{}}XCBS-A\end{tabular} 
		& No & Yes & 4  & 4 & 2& 4 \\ \hline
		\begin{tabular}[c]{@{}c@{}}Priority Based\end{tabular} 
		& No & Yes & 2 & 1 & 4& 1 \\ \hline
		\begin{tabular}[c]{@{}c@{}}\textbf{DeMAPF}\end{tabular} 
		& \textbf{Yes} & \textbf{Yes} & \textbf{1}  & \textbf{2} & \textbf{3}& \textbf{2}\\ \hline
		
	\end{tabular}
\end{table*}

\section{Performance Evaluation and Discussion}\label{section_result}
\subsection{Test setup}
The grid-based maps~\cite{Sturtevant2012} and scenarios have been released as MAPF benchmark sets. The empty-48-48.map from the benchmark was used for most of the experiments described here. The maps were modified to define edges of uniform length and the scenarios were modified to include agent characteristics like length and speed of the Traveller agents. The performance of DeMAPF was compared by running the testcases on the executables provided by the respective authors for PDDL based approach~\cite{Khemani2012}, Extended Conflict Based Search (XCBS)~\cite{Thomas2015},  XCBS-A~\cite{Thomas2018}. The authors also implemented a Priority based approach along the lines of DiMAPP~\cite{Niyogi2015} for evaluation. DeMAPF is implemented using Java8 and all the tests were run on an Intel Xeon(R) Silver CPU with 48 cores and 15GB RAM. \\

\subsection{Results}
\textbf{Number of Traveller Agents:} In Figure~\ref{fig:Performance-grid}, we evaluated the time taken to arrive at a solution as the number of Travellers increases. For empty-48-48 map, it is observed that as the number of Travellers increase DeMAPF arrives at solution faster than most of the other algorithms. The PDDL based solution did not give results for all the agents as the planner LPG-td2.0~\cite{Gerevini2004} failed to scale up to meet the number of facts instantiated.\\

\textbf{Different map sizes}: Figure~\ref{fig:networks} compares the time taken arriving at solution for different types of graphs. Here multiple maps from the MAPF benchmark were used as shown in the figure and the number of Traveller agents were kept constant at 32 (corresponding to the number of agents in the smallest map). It is observed that for larger maps, the number of Travellers agents being less led to lesser conflicts, hence faster convergence. Even in these scenarios, it is seen DeMAPF performs consistently better than the other distributed approaches.\\ 

\textbf{Memory Utilisation}: We evaluated the amount of memory used by each of the algorithms. This was obtained by calculating the runtime memory available during the execution of the algorithms. All the algorithms were run on the same machine for this evaluation on the empty-48-48.map. Figure~\ref{fig:Memory} shows that the other distributed approaches have significantly higher memory consumption which can be explained because of a centralised agent for maintaining the search tree of plans of all agents. On the other hand, DeMAPF has a smaller memory foot-print as because the agents plan for their solutions independently with smaller search trees. \\

\textbf{Solution Quality}: The Figure~\ref{fig:SolutionQuality} shows the difference in the solution cost arrived to by the multiple approaches for the empty-48-48.map. XCBS which is proven to be an optimal approach was chosen as benchmark and the total solution cost, was measured against the solution cost of the XCBS solution. The figure shows the inflation in solution cost over the optimal values. XCBSA, Priority based approach and DeMAPF all lead to suboptimal solutions. However it should be noted that the cost of solution in case of priority based approach was more than the DeMAPF solution. \\

\textbf{Distributivity}: DeMAPF is a distributed algorithm, and we checked the scalability of the algorithm by uniformly distributing the agents over multiple machines. The agents are implemented as JADE~\cite{Jade2005} agents, which is a popular standard multi agent development framework. The JADE environment was running commonly across all machines, which enabled the seamless communication between the agents. In Figure~\ref{fig:Distributivity}, the 'x' in the legend 'DeMAPF-x' indicates the number of machines on which the algorithm was simultaneously executed for the empty-48-48.map for upto  100 Travellers. During the run, the Travellers were started on different machines randomly but ensuring equitable distribution. It is observed that as expected the performance of the algorithm improved as the agents were distributed over multiple machines. It should be noted that the performance improvement was seen significantly when the number of agents planned for was large. The time taken is also dependent on the underlying JADE framework which was used for the exchange of messages between agents. An effective middleware or messaging broker for communication may bring down the processing times significantly.\\

\indent Table~\ref{table-summary} summarises the ranking of the different algorithms in terms of the parameters they were evaluated against. It is seen that DeMAPF stands better than all the other algorithms and demonstrates lower consumption of memory with faster convergence to solutions. 

\section{Future Work}\label{section_futurework}
\indent The approach has a few limitations which are proposed to be worked upon. First, the precedence ordering of the Travellers restricts an optimal allocation of the network resources by the Routers, hence a better heuristic for the allocation by the Routers has to be looked into. Second, in some scenarios a higher precedence Traveller may not accept a proposed allocation by a Router and look for alternative plans, making a favourable allocation by the same Router, for a lower precedence Traveller possible. This scenario is not handled in the current proposal and efforts have to be made to address this issue in the future. Finally, this effort assumes complete and unfettered communication between agents, future efforts will have to be directed towards handling irregular, lossy communication networks as well.\\

 \section{Conclusion} \label{section_conclusion}
\indent In this paper, we have a proposed a novel decentralised approach to handle MAPF for spatially extended agents. The proposed approach, Decentralised Multi Agent Path Finding (DeMAPF), generates conflict free plans for the Travellers by allocating the road space, in the order of precedence of the Travellers, as per their computed best possible plan. The decisions are made by the agents individually, based on messages received by it. No single agent maintains a full plan of all the agents at any time, thereby making it a decentralised solution. This is necessary in scenarios where the privacy of the agent has to be maintained. Secondly, in terms of the compute resources, as each agent handles its own computation, the requirement of compute resources in terms of processing and memory is distributed, which enables the solution to be deployed  and executed in a distributed manner, on several resource constrained platforms as well.\\
\indent DeMAPF gives a decentralised approach to handling MAPF, while reducing compute requirement, enabling distributivity and demonstrating scalability to handle large scale problems. This approach is novel in terms of the mechanism by which the decentralisation of the problem has been attempted, every Router agent manages only one entity, every Traveller agent only plans for itself. A uniform precedence ordering of the Traveller across all Routers ensures that the solution will converge to a conflict-free solution. The performance of DeMAPF when evaluated against several other distributed approaches has been better than most of the existing approaches.

\bibliographystyle{named}

\bibliography{draft7}

\end{document}